\documentclass[12pt,preprint]{aastex}
\usepackage{amsmath}
\usepackage{graphicx}
\usepackage{color}
\usepackage{url}
\usepackage{CJKutf8}
\usepackage{wasysym}

\slugcomment{Accepted in \aj~2016}

\shorttitle{332P}
\shortauthors{Hui et al.}

\begin{document}

\title{Constraints on Comet 332P/Ikeya-Murakami}
\author{
\begin{CJK}{UTF8}{bsmi}
Man-To Hui (許文韜)$^{1}$, Quan-Zhi Ye (葉泉志)$^{2}$, 
\end{CJK}
and Paul Wiegert$^{2,3}$
}
\affil{$^1$Department of Earth, Planetary and Space Sciences,
UCLA, 
595 Charles Young Drive East, 
Los Angeles, CA 90095-1567\\
}
\affil{$^2$Department of Physics and Astronomy, 
The University of Western Ontario,
London, Ontario, N6A 5B8, Canada\\
}
\affil{$^3$Centre for Planetary Science and Exploration, 
The University of Western Ontario, 
London, Ontario N6A 5B8, Canada\\
}
\email{pachacoti@ucla.edu}

\begin{abstract}
Encke-type comet 332P/Ikeya-Murakami is experiencing cascading fragmentation events during its 2016 apparition. It is likely the first splitting Encke-type comet ever being observed. A nongravitational solution to the astrometry reveals a statistical detection of the radial and transverse nongravitational parameters, $A_{1} = \left(1.54 \pm 0.39\right) \times 10^{-8}$ AU day$^{-2}$, and $A_{2} = \left(7.19 \pm 1.92\right) \times 10^{-9}$ AU day$^{-2}$, respectively, which implies a nucleus erosion rate of $\left(9.1 \pm 1.7\right)$\permil~per orbital revolution. The mass-loss rate likely has to be supported by a much larger fraction of an active surface area than known cases of short-period comets; it may be relevant to the ongoing fragmentation. We failed to detect any serendipitous pre-discovery observations of the comet in archival data from major sky surveys, whereby we infer that 332P used to be largely inactive, and is perhaps among the few short-period comets which have been reactivated from weakly active or dormant states. We therefore constrain an upper limit to the nucleus size as $2.0 \pm 0.2$ km in radius. A search for small bodies in similar orbits to that of 332P reveals comet P/2010 B2 (WISE) as the best candidate. From an empirical generalised Jupiter-family (Encke-type included) comet population model, we estimate the likelihood of chance alignment of the 332P--P/2010 B2 pair to be 1 in 33, a small number indicative of a genetic linkage between the two comets on a statistical basis. The pair possibly originated from a common progenitor which underwent a disintegration event well before the twentieth century.

\end{abstract}

\keywords{
comets: general --- comets: individual (332P/Ikeya-Murakami, P/2010 B2 (WISE)) --- methods: data analysis
}

\section{INTRODUCTION}
Comets are conceived to be amongst the most pristine objects originated from the Oort and the Kuiper reservoirs, and the scattered comet disc in the solar system (e.g., Dones et al. 2004; Ducan et al. 2004). Historically, long and short period comets are distinguished by an orbital period cutoff line arbitrarily set to be 200 years, where the former likely came from the Oort Cloud and the latter are more related to the Kuiper Belt and the scattered comet disc. A more appropriate taxonomic scheme to classify comets by Levison (1996) is to use $T_{\mathrm{J}}$, the Tisserand invariant with respect to Jupiter, defined by

\begin{equation}
T_{\mathrm{J}} = \frac{a_{\mathrm{J}}}{a} + 2 \sqrt{\frac{a}{a_{\mathrm{J}}}  \left(1 - e^{2}\right) } \left[ \cos i \cos i_{\mathrm{J}} + \sin i \sin i_{\mathrm{J}} \cos \left(\Omega - \Omega_{\mathrm{J}} \right) \right]
\label{eq_TJ},
\end{equation}

\noindent where $a$, $e$, $i$, and $\Omega$ are respectively the comet's semimajor axis, eccentricity, inclination, and longitude of ascending node, and the orbital elements with the subscript J refer to the Jupiter's. Comet 332P/Ikeya-Murakami (hereafter 332P) has $T_{\mathrm{J}} = 3.01$ and $a = 3.09$ AU, matching the definition of an Encke-type comet by Levison (1996). 

332P remained undiscovered until early 2010 November, when two Japanese amateur astronomers, K. Ikeya and S. Murakami, visually detected it at a magnitude of $\sim$9 (Nakano \& Ikeya 2010). The later change in morphology and evolution of the brightness suggest that the comet had experienced an outburst possibly triggered by crystallisation of buried amorphous water ice (Ishiguro et al. 2014). The comet was discovered falling apart into more than 30 pieces in its 2016 return, and the cascading fragmentation is still happening as time evolves (Kleyna et al. 2016; Jewitt et al., in preparation).

While it is not rare to witness comets suffering from disintegration events (e.g. Chen \& Jewitt 1994; Levison et al. 2002; Belton 2015), 332P is likely the first Encke-type comet which is being observed to break up.\footnote{Following a scheme by Jewitt et al. (2015), we do not count fragmented comets such as P/2013 R3 (Catalina-PANSTARRS) and P/2016 J1 (PANSTARRS) as Encke-type members, but active asteroids (a.k.a. main-belt comets).} Therefore, we feel the necessity to study this comet in more depth and present our analysis in this paper, for sake of our better understanding of evolution of comets in the solar system.

\section{ORBIT DETERMINATION}
\label{sec_orbdet}
Astrometric data of 332P were collected from the Minor Planet Electronic Circulars (MPEC) of the Minor Planet Center (MPC) and from Kleyna et al. (2016), who made use of the Panoramic Survey Telescope and Rapid Response System (Pan-STARRS) and the Canada France Hawaii Telescope (CFHT) telescopes.\footnote{We obtained 971 available astrometric observations in total as of 2016 May 04.} We attempted to establish an orbit linkage between 502 observations of the component C of 332P from apparition 2016 and 469 observations from 2010-2011, when none of the observations suggest detection of the breakup, as the component C is the primary of the nucleus (Dekelver \& Sekanina 2016; Kleyna et al. 2016). The observations were debiased and weighted based upon the scheme by Chesley et al. (2010) and Farnocchia et al. (2015). For CFHT observations, we increased the uncertainties based on seeing and morphology of the comet. Weightings of the remaining observations were further degraded by a factor of two, whereby we think that it will incorporate centroiding errors of the comet, since we have no knowledge about their observations details. The final weighting scheme is a conservative estimate of the uncertainties. The orbit determination was done with \textit{EXORB8} code\footnote{The code is a part of the \textit{SOLEX} package, which is available at \url{http://www.solexorb.it/Solex120/Download.html}.}, which utilises the JPL DE431 ephemeris, includes perturbations from the eight major planets, Pluto, and the most massive 27 asteroids in the main-belt region, as well as post-Newtonian relativistic effects due to the Sun, although the influences are negligible. During preliminary orbital computations, observations with astrometric residuals $>$2\arcsec.0 from \textit{ad hoc} osculating solutions were discarded iteratively. The threshold was chosen as a compromise between exclusion of bad outliers and keeping as many data as possible. We obtained orbital solutions with and without solving for $A_{1}$ and $A_{2}$, respectively the radial and transverse nongravitational parameters, which come from a nongravitational force model symmetric about perihelion, with the assumption that the nongravitational acceleration is proportional to the sublimation rate of water-ice (Marsden et al. 1973). Normally nongravitational parameters are reliably obtainable if there are astrometric observations covering more than three apparitions (Yeomans et al. 2004), unless the number of measurements is sufficiently large (e.g., Marsden 1997; Farnocchia et al. 2014). A zero normal nongravitational parameter, $A_{3} = 0$, is assumed, since during our preliminary test, solving for it was found to have no improvement to the best fit, and there is no statistically meaningful detection of it either, which is in agreement with Yeomans et al. (2004) that $A_{3}$ is normally orders of magnitude smaller than the radial and transverse terms. In order to test the reliability of the nongravitational parameters, we ran Monte Carlo selections of the astrometric observations in a stochastically varying fraction sense, and recomputed best-fit orbital solutions in \textit{EXORB8} for 2,000 iterations. During each run 30--70\% of observations were randomly selected for orbit determination. We then calculated mean and standard deviation values of the nongravitational parameters. 

We summarise the solutions to the orbit of 332P in Table \ref{tab_orb}, from which we can see that the two methods give consistent nongravitational parameters, both with significant detection confidence. It is noteworthy that the values are apparently different from those given by JPL HORIZONS,\footnote{Retrieved on 2016 May 04, the detailed values of the nongravitational parameters are $A_{1} = \left( 6.68 \pm 0.71 \right) \times 10^{-8}$ AU day$^{-2}$, and $A_{2} = \left(2.62 \pm 0.26 \right) \times 10^{-8}$ AU day$^{-2}$ calculated by the JPL HORIZONS.} as well as those by Kleyna et al. (2016). The major difference between the JPL HORIZONS and this work lies in our use of new high quality measurements from the CFHT. We have verified that removal of the CFHT data or assigning them a low weighting ($\gtrsim$2\arcsec) gives us results comparable to those by the JPL HORIZONS. On the other hand, the results computed by Kleyna et al. (2016) reflect the fact that there was a limited number of available astrometric measurements and the covered arc spanned too short a time, and therefore the calculation could have been easily influenced by potential low quality astrometry from other observers. 

\begin{figure}
  \centering
  \begin{tabular}[b]{@{}p{1\textwidth}@{}}
    \centering\includegraphics[scale=0.81]{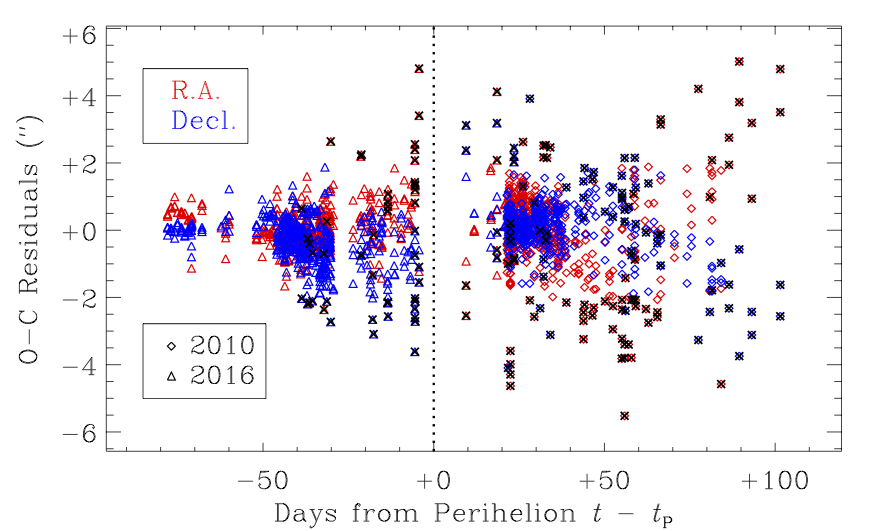} \\
    \centering\small(a)
  \end{tabular}%
  \quad
  \begin{tabular}[b]{@{}p{1\textwidth}@{}}
    \centering\includegraphics[scale=0.81]{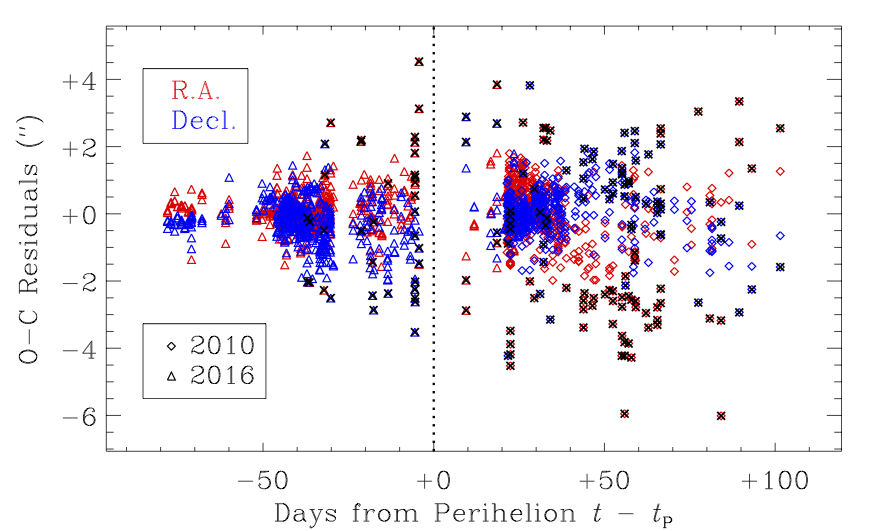} \\
    \centering\small (b)
  \end{tabular}
  \caption{
Plots of O$-$C residuals in right ascension (R.A.) and declination (Decl.) as functions of time in (a) the pure gravitational solution, and (b) the nongravitational solution. Nearest perihelion epochs are used as reference time. Residuals in R.A. and in Decl. are respectively coloured in red and blue. Diamonds and triangles label residuals from the 2010 and the 2016 apparitions, respectively. Crosses correspond to rejected observations. The nongravitational solution slightly improves the scatter of the data points, especially those from the 2010 apparition.
\label{fig_res}
  }
\end{figure}

Figure \ref{fig_res} shows astrometric residuals (the differences between the observed and calculated positions, i.e., O$-$C residuals) as a function of time in the solutions without and with inclusion of the nongravitational parameters. Although the nongravitational solution fails to significantly improve the scatter of the data points, it does render us a smaller best-fit RMS while keeping more astrometric data unfiltered than does the pure gravitational solution (see Table \ref{tab_orb}). The ongoing fragmentation events morphologically show that the nucleus of 332P is suffering from nongravitational forces. Our confident detection of the nongravitational parameters does confirm the existence of the nongravitational effect astrometrically. 

It is noteworthy that we notice a tiny systematic bias in the astrometric residuals starting from CFHT observations from UT 2016 April 09, which likely indicates that the nongravitational force model is not perfect. The observed positions of the optocentre of component C are constantly southeast off from the calculated positions by $\sim$0\arcsec.7 on UT 2016 April 09 and even $\sim$1\arcsec.0 on April 13, which can be removed by neither of the orbital solutions. We have ruled out the possibility that it is due to unnoticed biases in star catalogues or timing errors, because another fragment of the comet, component A, which is isolated from any other debris clouds or components, shows no evidence of such a systematic bias in its astrometric residuals. The centroiding error is $\lesssim$0\arcsec.2 (Micheli, private communication), too small to account for the offset. We inspected all the CFHT images and found that, intriguingly, a protrusion of component C of $\sim$1\arcsec.9 in length, elongated to the west, came into being starting from UT 2016 March 31, and that it was more readily seen in images from UT 2016 April 05. However, this feature was missing in images from UT 2016 April 09 but substituted with at least three new faint components, in spite of similar seeing. Unfortunately images from UT 2016 April 13 suffered from poor seeing and therefore none of the new components were seen. We thus favour that the systematic bias starting from UT 2016 April 09 is likely related to the newly developing fragmentation event, yet follow-up observations are needed to confirm this idea.

\section{ARCHIVAL DATA SEARCH}
\label{subsec_archive}

We searched for serendipitous pre-discovery detections of comet 332P in the survey pointing dataset available at the MPC\footnote{http://www.minorplanetcenter.net/iau/SkyCoverage.html}, which stores about five million pointing records by various sky surveys. Since we were aware of the faintness of the comet if it is around its aphelion, by assuming that the comet would be among the brightest when it is closer to both the Sun and the Earth, we focused on periods of time no longer than six months from perihelion passages of the comet in different years. The ephemeris of the comet was computed with the nongravitational orbital solution in Table \ref{tab_orb} with orbital epochs close to corresponding perihelia in different years. This also helps to overcome a shortcoming in our code for archival search that it currently cannot deal with nonzero nongravitational parameters. We have verified that during the selected time intervals, the differences in positions are well covered by the fields-of-view (FOVs) of the sky surveys listed at the MPC. In this way, a total of six pointing records in 1999 (two hits) and 2005 (four hits) which might include the comet were found. Amongst these data we were able to obtain eight raw images, taken respectively on UT 2005 April 19 and 30 by the Catalina Sky Survey (CSS), all in 30s exposures. The 3$\sigma$ uncertainty ellipse calculated from the covariance matrix of orbital elements is $\sim$0\degr.3 then, which is small enough compared to the field-of-view of CSS images ($2\degr.9 \times 2\degr.8$). We actually extended our search for 332P far beyond the nominal positions and the uncertainty ellipse and the whole FOVs of the images were scrutinised. However, unfortunately, we still failed to discover any uncategorised moving objects in the images from the two days. The limiting magnitude of the CSS images is 19.5, corresponding to a signal level above 3$\sigma$ of the background noise.

Additional search was performed by querying the Solar System Object Image Search (Gwyn et al. 2012) page at the Canadian Astronomical Data Centre (CADC). As the CADC archives data from large telescopes such as the CFHT, and thereby limiting magnitudes in these images are much fainter, we decided not to apply the aforementioned constraint on time which we used to search for the survey pointing dataset. A number of images taken on UT 2003 September 25 and 27 at the CFHT as part of the CFHT Legacy Survey were found at or within 1\degr~of the expected position of 332P in 2003. These are 70s MegaCam ($\sim 1\degr \times 1\degr$ FOV) exposures taken in the Sloan-g' filter. At this time, 332P would have been $\sim$4 AU from the Sun and with a JPL Horizons predicted nuclear magnitude of 18.0. However we expect that the actual apparent magnitude could be much fainter.

The images were searched for moving objects using a pipeline previously used successfully to search for main-belt comets and asteroids (Wiegert et al. 2007; Gilbert \& Wiegert 2009, 2010; August \& Wiegert 2013). The pipeline extracts image sources using the Source Extractor package (Bertin \& Arnouts 1996), and examines the source list and reports on pairs or triplets of sources consistent with a solar system body moving within some user-defined parameter range.

There were four sets of images from 2003 September 25 searched, each consisting of two images of the same field taken 1.3--1.5 hours apart. These were examined for moving objects consistent with the expected sky rate of motion of 332P, down to 1.5$\sigma$ above the image background (limiting magnitude 23.1). No candidates that could be successful linked with earlier observations of 332P were found.

Three sets of three images taken approximately 1.3 hours apart on 2003 September 27 were also searched without success. One of these sets, which covered the bulk of the uncertainty region of 332P on the sky was re-examined including all sources down to 1.0$\sigma$ above the image background (limiting magnitude $\sim$23.6), but no suitable moving sources were found. Unfortunately, data for the chip (MegaCam chip 3) on which 332P was nominally expected on September 27 was blank. This was true for all images we downloaded in this time frame, and so may reflect a technical issue at the telescope at the time. In any case, whether 332P was too faint, not in the images, or fell into a gap or near a bright star, we were unable to find any moving sources consistent with 332P's expected motion.

\section{DISCUSSIONS}

\subsection{Nucleus Erosion Inferred from Nongravitational Parameters}
\label{sec_mloss}
The significant detection of the nongravitational parameters $A_{1}$ and $A_{2}$ indicates a nongravitational force acting on the nucleus, which is caused by anisotropic outgassing activities in most cases. Using the same method described in Hui et al. (2015), we can estimate the nucleus erosion $\eta$ based upon conservation of momentum by

\begin{align}
\nonumber
\eta \left(t_{1}, t_{2} \right) & \triangleq 1 - \frac{\mathcal{M} \left( t_{2} \right) }{\mathcal{M} \left( t_{1} \right)}\\
\label{eq_eros_ng}
 & = 1 - \exp \left[-\frac{\sqrt{A_{1}^{2} + A_{2}^{2} + A_{3}^{2}} }{\kappa} \int_{t_{1}}^{t_{2}} \frac{g \left(r \left( t \right) \right)}{v \left( t \right)} \mathrm{d}t \right] \\
 & \simeq \frac{\sqrt{A_{1}^{2} + A_{2}^{2} + A_{3}^{2}} }{\kappa} \int_{t_{1}}^{t_{2}} \frac{g \left(r \left( t \right) \right)}{v \left( t \right)} \mathrm{d}t
 \label{eq_eros_ng2} ,
\end{align}

\noindent where $\eta \left(t_{1}, t_{2} \right)$ denotes the nucleus erosion between time $t_{1}$ and $t_{2}$, $\mathcal{M}$ is the nucleus mass, $r$ is the heliocentric distance of the nucleus, $v$ is the outflow speed, $\kappa$ is a dimensionless collimation efficiency in the range $0 \le \kappa \le 1$, the two boundaries corresponding for isotropic emission and perfectly collimated ejection respectively, and $g \left( r \right)$ is a standard momentum-transfer empirical law of water-ice sublimation as a function of heliocentric distance introduced in Marsden et al. (1973). Equation (\ref{eq_eros_ng2}) is normally a good estimate, as the exponential power in Equation (\ref{eq_eros_ng}) is generally minute.

We approximate the outflow speed by mean thermal speed $v_{\mathrm{th}} = \sqrt{8 k_{\mathrm{B}} T / \left( \pi \mu \right)}$. Here $k_{\mathrm{B}} = 1.3806 \times 10^{-23}$ J K$^{-1}$ is the Boltzmann constant, $\mu$ is the molecular mass, and $T$ is the surface temperature of the nucleus, which can be numerically solved from the following equation due to energy conservation

\begin{equation}
\frac{\left(1 - \mathrm{A_{B}} \right) S_{\odot}}{r^{2}} \cos \zeta = \epsilon \sigma T^{4} + L\left( T \right) Z\left( T \right)
\label{eq_sublim},
\end{equation}

\noindent where $\mathrm{A_{B}}$ is the Bond albedo, $S_{\odot} = 1361$ W m$^{-2}$ is the solar constant, $\cos \zeta$ is the effective projection factor for the surface, $\epsilon$ is the emissivity, $\sigma = 5.6704 \times 10^{-8}$ W m$^{-2}$ K$^{-4}$ is the Stefan-Boltzmann constant, $L$ is the latent heat of water-ice sublimation, and $Z$ is the gas production rate on a unit area surface. We assume $\mathrm{A_{B}} = 0.01$, $\epsilon = 1$, $\kappa = 0.5$, and an isothermal nucleus, i.e., $\cos \zeta = 1/4$. The assumed parameters in Equation (\ref{eq_sublim}) within known ranges are found to have minimal effects on the obtained mean thermal speed.

We consider the erosion of the cometary nucleus during a complete orbit around the Sun and choose $t_{1} = 0$ to be the last epoch of perihelion passage as a reference. With the above values, Equation (\ref{eq_eros_ng2}) yields the erosion ratio per orbital revolution as large as $\eta \left(0, P \right) \simeq \left(9.1 \pm 1.7\right)\permil$, where $P$ is the orbital period of 332P. If the nucleus is $\sim$1 km in radius (Ishiguro et al. 2014), and has a typical bulk density 0.4 g cm$^{-3}$, we obtain a nucleus mass-loss rate of $89 \pm 16$ kg s$^{-1}$ due to the cometary activity alone, and the active surface area has to be roughly over twice the nucleus surface area in order to support such an amount of outgassing, which shows a strong contrast to some known cases of short-period comets, e.g., only $\sim$4\% for comet 67P/Churyumov-Gerasimenko (Combi et al. 2012). Taking the dusty nature of 332P into consideration, this is likely related to the ongoing fragmentation events of the comet. The mass-loss rate can be translated into a reduction rate of nucleus radius of $3.0 \pm 0.6$ m per orbital revolution, which means that the physical lifetime of the comet is $\sim$1 kyr, considerably shorter than the median dynamical lifetime of short-period comets (Levison \& Duncan 1994), by two orders of magnitude. 



\subsection{Constraint on Nucleus Size}
The non-detection of 332P in archival data from CSS in 2005 can provide us with a constraint on its nucleus size. We summarise the observing conditions of the comet in Table \ref{tab_geo}. The relationship between the apparent magnitude of the cometary nucleus $m$ and the effective geometric cross-section of the nucleus $C_{\mathrm{e}}$ is given by

\begin{equation}
C_{\mathrm{e}} = \frac{\pi r^{2} \mathit{\Delta}^{2}}{\phi p_{\lambda} r_{\oplus}^{2}} 10^{-0.4 \left(m_{\lambda} - m_{\odot, \lambda} \right)}
\label{eq_Ce},
\end{equation}

\noindent where $\mathit{\Delta}$ is the distance between the comet and the observatory, $r_{\oplus} = 1$ AU is the mean distance between the Sun and the Earth, $p_{\lambda}$ is the geometric albedo at some wavelength $\lambda$, $\phi$ is the $HG$ photometric system phase function developed by Bowell et al. (1989), and $m_{\odot}$ is the apparent magnitude of the Sun. Unfortunately the archival images were taken without any photometry-standard filter, and thus, we have to approximate the bandpass as $V$-band. We determine the limiting magnitude of the CSS images as the magnitude corresponding to $3\sigma$ of the background level, thereby yielding $m_{V} > 19.5$ for the nucleus of 332P. Assuming a nominal albedo of $p_{V} = 0.04$ and a typical slope parameter $G = 0.15_{-0.10}^{+0.08}$ (c.f. Lagerkvist \& Magnusson 1990), we yield $C_{\mathrm{e}} < 13.1_{-2.2}^{+3.0}$ km$^{2}$ as the constraint on the effective cross-section of the nucleus, corresponding to a nucleus radius $R_{\mathrm{N}} < 2.0 \pm 0.2$ km, which is in exact agreement with Ishiguro et al. (2014).

We can also set an upper limit to the active area of the nucleus from the non-detection in the CSS data. Using the empirical relationship by Jorda et al. (2008), the water production rate of 332P during the Catalina observations can be converted from the limiting magnitude as $Q \lesssim 1.3 \times 10^{26}$ molecules s$^{-1}$. Equation (\ref{eq_sublim}) yields $Z \simeq 8 \times 10^{20}$ molecules s$^{-1}$ m$^{-2}$ for $r \simeq 1.6$ AU, given that the nucleus is isothermal. Therefore, the active area of the nucleus would be $\lesssim$0.17 km$^{2}$, smaller than the upper limit of the effective nucleus cross-section by two orders of magnitude. Note that in reality it is impractical that a cometary nucleus is isothermal. By this token, the active area should be even smaller than our estimate. We therefore conclude that comet 332P very likely used to be largely inactive, thereby preventing its discovery in previous apparitions before 2010. It is likely among the few cases where a short-period comet is identified to have been reactivated from a previous dormant or weakly active state.

Provided that the non-detection in the CFHT data is due to the faintness of the comet as well, the size constraint on the nucleus is even stricter. By assuming $p_\mathrm{g} = p_V = 0.04$, and repeating completely the same steps as above, we obtain $C_\mathrm{e} \lesssim 0.7$ km$^{2}$ for its effective cross-section from Equation (\ref{eq_Ce}), or equivalently $R_\mathrm{N} \lesssim 0.5$ km for the nucleus radius. But the non-detection in the CFHT data does not provide a stricter constraint on the active area of the nucleus.

\subsection{Evolution of the Orbit}
\label{sec_orbevol}
Ishiguro et al. (2014) suggested that 332P was likely a Jupiter-family member originating from the Kuiper belt. However, since their work was based upon an orbital solution with observations covering merely 80 days, and now uncertainties in our orbital elements significantly reduce thanks to the much longer arc, we feel the necessity to reinvestigate its orbital evolution. Our method is similar to the one by Ishiguro et al. (2014).

We exploit \textit{EXORB8} to generate 200 clones of 332P whose initial orbital elements are synthesised from the covariance matrix of orbital elements computed during our orbit determination. The output clones are then fed into \textit{SOLEX} for orbital integration backward to the past 1 kyr and forward to the next 1 kyr separately. Configurations of perturbations are set completely the same as described in Section \ref{sec_orbdet}. We have to assume that the nongravitational parameters remain constant, which, in some cases unfortunately, is a poor approximation (e.g., Sekanina 1972; Yeomans et al. 2004). But we expect that the clones with different nongravitational parameters should provide us with some hints about different possibilities of orbital evolution of 332P.

\begin{figure}
\epsscale{1.0}
\begin{center}
\plotone{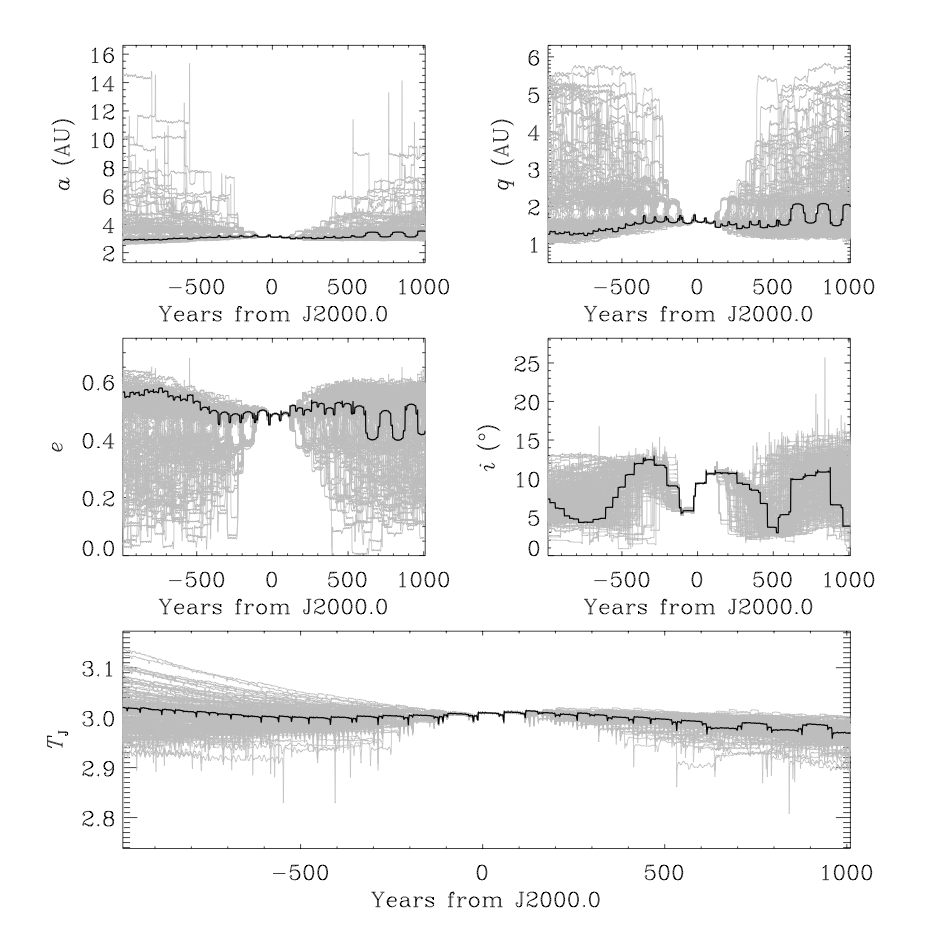}
\caption{Orbital evolution of the nominal orbit (black) and the 200 Monte Carlo clones (grey) of 332P/Ikeya-Murakami from a millennium in the past to the next millennium in the future. The year here refers to the Julian year. The nongravitational parameters of each clone are all assumed to remain unchanged during the period of time.
\label{fig_osc_V1}
} 
\end{center} 
\end{figure}

The results are shown in Figure \ref{fig_osc_V1}. We notice that the orbit of 332P likely has been chaotic for the past 1 kyr, and may continue the status for the next 1 kyr. Provided that our assumption about its constant nongravitational parameters holds, auxiliary simulations we have run indicate that its current Lyapunov time is $\sim$25 yr. From now on a secular decline in its Tisserand invariant with respect to Jupiter $T_{\mathrm{J}}$ seems inevitable, thereby turning 332P into a JFC (Jupiter-family comet) within the next 1 kyr. On the other hand, we are unable to identify the origin of the comet, in that $T_{\mathrm{J}} > 3$ and $< 3$ seems equally possible in the past, based upon our simulation. This reinforces the idea that Jupiter-family and Encke-type comets can be mutually related.

\begin{figure}
\epsscale{1.0}
\begin{center}
\plotone{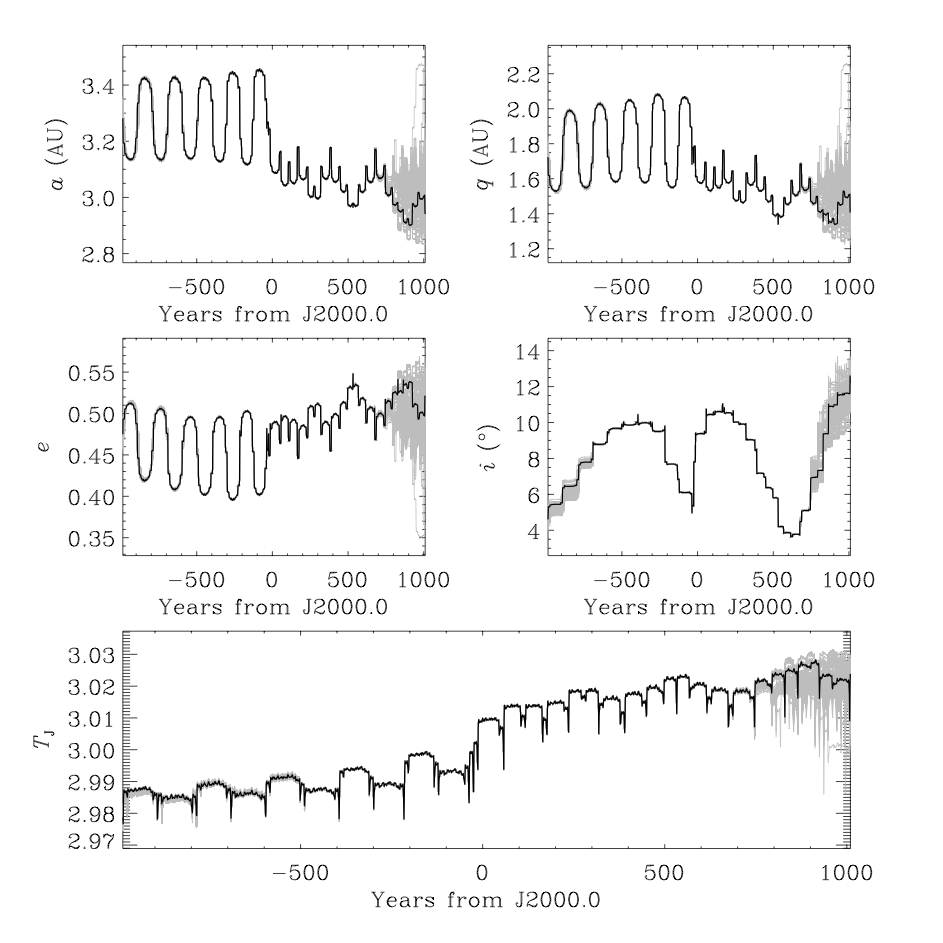}
\caption{Orbital evolution of the nominal orbit (black) and the 200 Monte Carlo clones (grey) of 332P/Ikeya-Murakami from a millennium in the past to the next millennium in the future, without inclusion of the nongravitational parameters. The year here refers to the Julian year. The strong dispersion of clones shown in Figure \ref{fig_osc_V1} is no longer present in this scenario.
\label{fig_osc_V1_pg}
} 
\end{center} 
\end{figure}

Intriguingly, if we integrate orbits of the clones generated from the pure gravitational solution over the same period of time, the dispersion of the clones appears distinctly reduced (see Figure \ref{fig_osc_V1_pg}). We calculate their Lyapunov time to be only slightly longer than in the nongravitational case, $\sim$50 yr, but in this case the clones may all be within a more weakly chaotic region. The past 1 kyr may have witnessed oscillations in its semimajor axis, perihelion distance and eccentricity, with an oscillatory period $\sim$190 yr, under a weak influence of the 2:1 Jovian resonance, since the resonant angle $\varphi = 2 \ell_{\mathrm{J}} - \ell - \varpi$, where $\ell$ and $\varpi$ are respectively the mean longitude and the longitude of perihelion, and the subscript $\mathrm{J}$ refers to Jupiter, varies nearly circularly, within a range of $\sim\pm170$\degr. If there were no nongravitational force exerting on 332P, it was likely expelled from the 2:1 Jovian resonance region very recently (within a century). The trend of its Tisserand invariant with respect to Jupiter would indicate a Jupiter-family origin, and may continue the current status as an Encke-type comet at least in the next 1 kyr, which is in agreement with Ishiguro et al. (2014). However, we do not favour this idea that comet 332P has no influence from the nongravitational effect, because this requires a perfectly isotropic outgassing scenario, which in practice can never be achieved.  Thus, we think that Figure \ref{fig_osc_V1} should be more representative about the actual dynamical evolution of 332P.

\subsection{Earlier Fragment Search}
\label{sec_search_frag}
We employ the following criteria to narrow our search for small bodies moving in orbits similar to 332P. The first step is to utilise the JPL Small-Body Database Search Engine\footnote{\url{http://ssd.jpl.nasa.gov/sbdb_query.cgi}} and set maximum tolerances around 332P's orbital elements. We choose $\Delta a \simeq \pm 0.5$ AU, $\Delta e \simeq \pm 0.05$, $\Delta q \simeq \pm 0.2$ AU, $\Delta i \simeq \pm 10$\degr, $\Delta \Omega = \Delta \omega \simeq \pm30$\degr, where $q$ and $\omega$ are perihelion distance and argument of perihelion, respectively, which are set roughly according to standard deviations in orbital elements of members in comet groups such as Kreutz, Meyer and Marsden groups. 

The second constraint is to exploit the D-criterion, which is a quantity parametrised by a five-dimensional phase space constructed with orbital elements of two different objects, in the following formalism
\begin{align}
\nonumber
D_{\mathrm{SH}}^{2} = &  \left(e_{1} - e_{2}\right)^{2} + \left(q_{1} - q_{2} \right)^{2} + 4 \sin^{2} \left(\frac{i_{1} - i_{2}}{2} \right) + 4 \sin i_{1} \sin i_{2} \sin^{2} \left( \frac{\Omega_{1} - \Omega_{2}}{2}\right) \\
 & + \left(e_{1} + e_{2} \right)^{2} \sin^{2} \left[\frac{ \omega_{1} -  \omega_{2}  }{2} + 
 \arcsin \left( \cos \frac{i_{1} + i_{2}}{2} \sin \frac{\Omega_{1} - \Omega_{2}}{2} \sec I_{12} \right) \right],
\end{align}
\noindent where $I$ can be calculated from
\begin{equation}
\sin^{2} I_{12} = \sin^{2} \left( \frac{i_{1} - i_{2}}{2} \right)+ \sin i_{1} \sin i_{2} \sin^{2} \left( \frac{\Omega_{1} - \Omega_{2}}{2}\right),
\end{equation}
\noindent and the subscripts differentiate the two orbits to be compared (Southworth \& Hawkins 1963). Generally the smaller $D_{\mathrm{SH}}$ is, the more similar the two orbits are. We somewhat arbitrarily set $D_{\mathrm{SH}} = 0.2$ as the maximum tolerance of the D-criterion. The search is done by scanning the MPC Orbit (MPCORB) Database. Note that all the orbital element comparisons were done at the same epoch. We did notice slight differences in orbital elements between the MPCORB file and JPL HORIZONS even if epochs are the same. However, these differences along with whether integrating orbits to the same epoch or not have been verified to have minimal influences on the D-criterion. 

Thirdly, we filter out targets with orbital uncertainty parameters, $U > 7$, defined by the MPC.\footnote{See \url{http://www.minorplanetcenter.net/iau/info/UValue.html} for the definition.} Thereby three candidates are obtained, which are summarised in Table \ref{tab_simorb}. We immediately notice that comet P/2010 B2 (WISE) (hereafter 2010 B2) is obviously the best candidate of all, as it has a nearly identical orbit to that of 332P, except a mere $\sim$3\degr~difference in its longitude of ascending node $\Omega$. Its nucleus size is constrained well by observations from the Wide-field Infrared Survey Explorer (WISE) mission, $0.5 \pm 0.1$ km in radius (Bauer et al. 2015), smaller than or possibly comparable to the size of 332P's nucleus.

\begin{figure}
\epsscale{1.0}
\begin{center}
\plotone{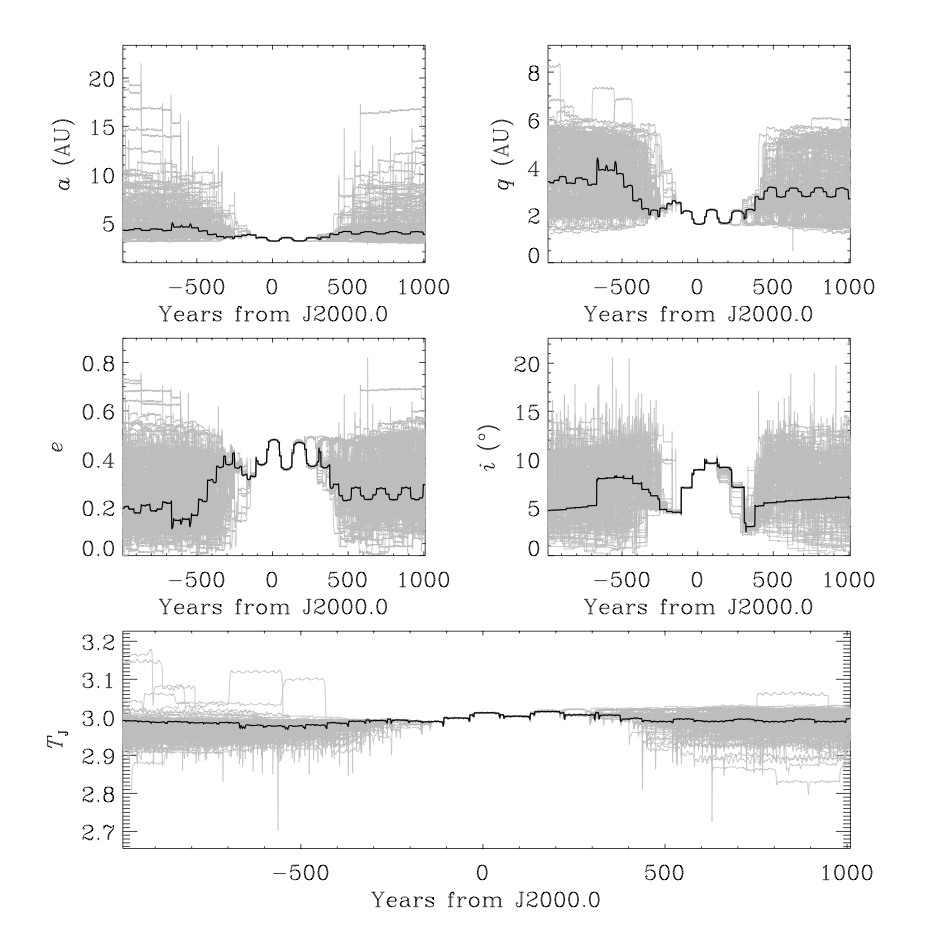}
\caption{Orbital evolution of the nominal orbit (black) and the 200 Monte Carlo clones (grey) of P/2010 B2 (WISE) from a millennium in the past to the next millennium in the future. The year here refers to the Julian year. The clones are assumed to have zero nongravitational parameters.
\label{fig_osc_B2}
} 
\end{center} 
\end{figure}

We briefly investigate the orbital evolution of 2010 B2 in hope that, if it is indeed related to 332P, we shall see similarities in evolution of both orbits. Following the same technique described in Section \ref{sec_orbevol}, we first integrate the orbits of 200 clones of 2010 B2 over the same period of time and examined the orbital evolution, assuming no nongravitational effects, which is due to the fact that observations of the comet only cover a very short arc. The results are illustrated in Figure \ref{fig_osc_B2}. In spite of significantly larger uncertainties in the orbital solution of 2010 B2, striking similarities between evolution of the orbits of 2010 B2 and 332P can be immediately seen, thus possibly indicating a common origin of the pair. A slightly different trend in evolution of the Tisserand invariant of 2010 B2 from the one of 332P may be attributed to the zero nongravitational force assumption. 2010 B2 may likely have originated from the Jupiter family in the past, however, the possibility that it had been an Encke-type member for over the past 1 kyr cannot be fully ruled out either. The next 1 kyr is likely to witness it retaining $T_{\mathrm{J}} \simeq 3$.

The first method cannot provide us with definitive conclusion proving/disproving the relationship between 332P and 2010 B2. Therefore we adopt the second method, which is to investigate if the pair had close encounters with each other in the past. For efficiency we pick up 50 clones for each of the comets, thereby offering us 2,500 different combinations of mutual distances, which is conceived to be sufficiently ample and representative. Here we only focus on close encounters $\le 0.1$ AU. Such close encounters appear to occur frequently before the late eighteenth century. Seemingly promising, the closest distance between the clones is $\sim$0.002 AU in 1500s. However, the majority of relative velocities distribute in a range of $\gtrsim$1 km s$^{-1}$, unrealistically enormous to be considered as separation velocity during disintegration. Plus a huge scatter of the data thereby hints poor reliability of this method.

Dissatisfied, we proceed to exploit the third method. Although we are aware that there is an optimisation method devised by Sekanina (1978, 1982), based upon orbital mechanics of cometary dust motions, which solves split parameters including the separation velocity $\vec{V}_{\mathrm{s}}$ of the secondary from the primary in terms of radial, transverse and normal components, $V_{\mathrm{R}}$, $V_{\mathrm{T}}$, and $V_{\mathrm{N}}$, the split epoch $t_{\mathrm{s}}$, and the differential deceleration $\gamma$, in accord with the simulation results of the close encounters between clones of 332P and 2010 B2, it is very likely that the pair was produced beyond the past 100 years, during which the chaoticity of the orbits of the two comets hampers us from applying the optimisation method. In fact a cursory attempt was made, which shows us that there is no convergence if we force the split event to occur within the past century. Therefore, we instead employ a generalised JFC population model, which includes typical Jupiter-family and Encke-type members, and follow the technique described in Wiegert \& Brown (2004), to calculate how many pairs in this family can have $D_{\mathrm{SH}} \le 0.04$.

\begin{figure}
\includegraphics[width=0.6\textwidth]{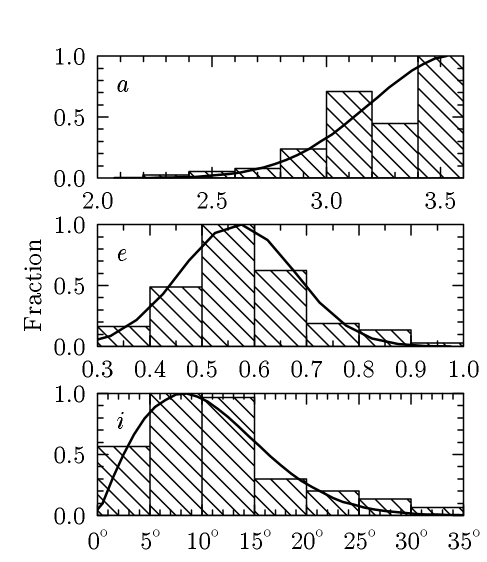}
\caption{Observed (bars) and modeled (lines) distributions of generalised JFCs near the orbits of 332P/Ikeya-Murakami and P/2010 B2 (WISE). We follow the philosophy in Grav et al. (2011) that, the distribution of $a$ is fitted using a Gaussian function centered at 3.6 AU with a width of 0.8 AU, the distribution of $e$ is fitted using a Gaussian function centered at 0.6 with a width of 0.2, and the distribution of $i$ is fitted with a Gaussian function with a width of $20$\degr~multiplied by $\sin{i}$.}
\label{fig:jfc_model}
\end{figure}

To estimate the expected number of such pairs, we need to know the true (debiased) distribution of the generalised JFC population. Unfortunately the demographics of generalised JFCs are poorly understood. However, the model by Grav et al. (2011, Fig. 24) hints that the detection of generalised JFCs with absolute (nuclear) magnitude $H < 20$ and semimajor axis $a < 3.5$~AU are largely completed. Therefore, we construct an \textit{ad~hoc} generalised JFC population model based upon the population of known generalised Jupiter-family members with $H < 20$ and $a < 3.5$~AU. We obtain a total of 97 known generalised JFCs that fall into this category as listed in the JPL Small-Body Database as of 2016 April 15. We fit the observed $a$, $e$ and $i$ following the procedure adopted by Grav et al. (2011) and generate a large number of generalised JFC population models (Figure~\ref{fig:jfc_model}). We assume uniform distributions for the secular orbital elements $\Omega$ and $\omega$. The true number of JFCs with $H < 20$ and $a < 3.5$~AU is loosely constrained. Fern{\'a}ndez et al. (2006) concludes that there are $N \sim 10^{3}$ JFCs with perihelion distance $q < 1.3$ AU and absolute magnitude $H < 22$; Grav et al. (2011) derives that there are $\sim$53,000 JFCs with $H < 24$. We therefore take $N \sim 10^{4}$ as an extremely conservative upper limit for our calculation. Consequently, we derive a probability of only 1 in 33 that a better match than the 332P--2010 B2 pair will be found. The small number statistically implies that the pair of 332P and 2010 B2 is likely genetically related.


\section{SUMMARY}
We present analyses of 332P and summarise what we found as follows:

\begin{enumerate}
\item The ongoing cascading fragmentation events of 332P are likely not new to this Encke-type comet. We find that comet 2010 B2 has an orbit very similar to the orbit of 332P, with $D_{\mathrm{SH}} = 0.04$. Based upon our generalised JFC population model, there is a probability of only 1 in 33 that a better pair than 332P--2010 B2 exists, which suggests that the two comets are likely genetically related and are fragments from a common progenitor. The results about close encounters between the pair likely indicate that the disintegration event occurred well before the twentieth century.

\item We obtained the radial and transverse nongravitational parameters of 332P to be $A_{1} = \left(1.54 \pm 0.39 \right) \times 10^{-8}$ AU day$^{-2}$ and $A_{2} = \left(7.19 \pm 1.92 \right) \times 10^{-9}$ AU day$^{-2}$ from the best fit to all the available astrometry. The nongravitational solution helps improve the scatter of O$-$C residuals slightly. The Monte Carlo random exclusion of observations has verified the stability of the two parameters.

\item The fractional mass-loss rate of the nucleus of 332P is estimated to be $\left(9.1 \pm 1.7\right)$\permil~per orbital revolution, based upon the nongravitational force experienced by the comet. If the nucleus is $\sim$1 km in radius and has a bulk density of 0.4 g cm$^{-3}$, the mass-loss rate is $89 \pm 16$ kg s$^{-1}$, which means that the active surface area is over twice the nucleus surface area to support the current activity, much larger than some known cases of short-period comets. It is likely related to the fragmentation events being observed in the apparition of the comet in 2016.

\item If the activity level of 332P persists, its physical lifetime is only $\sim$1 kyr, which is shorter than the median dynamical lifetime of short-period comets by two orders of magnitude.

\item We analyse the dynamical evolution of 332P from the past millennium to the next millennium. The orbit appears rather unstable if it goes further beyond $\sim$100 years in the past or in the future. Although we fail in identifying where the comet originated, it seems more likely that the comet will become a JFC within the next millennium, which reinforces the idea that Encke-type and Jupiter-family comets can be mutually related.

\item We searched for potential pre-discovery images of 332P in archival data from 2000 to pre-2010, without positive finding. The non-detection in the CSS archival data in 2005 suggests that the comet used to be inactive or weakly active before the outburst in 2010. The comet is perhaps among the few which are identified to have been reactivated from weakly active or dormant states. We constrain the nucleus size to be $R_{\mathrm{N}} < 2.0 \pm 0.2$ km in radius. A stricter constraint can be set on the nucleus size as $R_\mathrm{N} \lesssim 0.5$ km if the failure in detecting the comet in the CFHT data taken in 2003 was due to its faintness.

\end{enumerate}

\acknowledgements
Thanks to the anonymous referee for the prompt review and helpful comments. We are heartedly indebted to Aldo Vitagliano, the author of the \textit{SOLEX} package, for his tremendous efforts in implementing the codes and valuable discussions. We particularly thank Jan Kleyna, Marco Micheli and Richard Wainscoat from the CFHT and the Pan-STARRS teams for permission to use their astrometric measurements, Eric Christensen from the Catalina Sky Survey for providing us with the archival CSS imagery, David Clark for helping and guiding us search for archival observations, as well as all observers who have submitted astrometry to the MPC. Discussions with Davide Farnocchia have benefited this study. M.-T. specially appreciates Shigeki Murakami, with whom a daylong meetup about comet hunting was held in Cairns, Australia, a day after the total solar eclipse in 2012, for his visual discovery of comet 332P, as well as his wonderful gifts, including books by Tsutomu Seki. M.-T. also thanks David Jewitt for his comments on the manuscript, and for his financial support through a NASA grant. Q.-Z. thanks Peter Brown for his support. This research used the facilities of the Canadian Astronomy Data Centre operated by the National Research Council of Canada with the support of the Canadian Space Agency.

\begin{deluxetable}{c|rc|rc|rc|c}
\tabletypesize{\footnotesize}
\rotate
\tablecaption{Orbit of 332P/Ikeya-Murakami
(Reference: Heliocentric Ecliptic J2000.0)
\label{tab_orb}}
\tablewidth{0pt}
\tablehead{ Orbital & 
\multicolumn{2}{|c}{Solution PG\tablenotemark{a}}  & 
\multicolumn{2}{c}{Solution NG$_{1}$\tablenotemark{b}} & 
\multicolumn{2}{c}{Solution NG$_{2}$\tablenotemark{c}} & \colhead{Units} \\
Element & \colhead{Value} & \colhead{1$\sigma$ Uncertainty} & \colhead{Value} & \colhead{1$\sigma$ Uncertainty} & \colhead{Value} & \colhead{1$\sigma$ Uncertainty} &
}
\startdata
$e$ & 0.48907095 & 4.151$\times$10$^{-7}$ 
       & 0.48902071 & 6.189$\times$10$^{-6}$ 
       & 0.48902060 & 4.189$\times$10$^{-6}$ & -- \\ 
$a$ & 3.09041292 & 1.718$\times$10$^{-7}$ 
       & 3.09010201 & 4.126$\times$10$^{-5}$ 
       & 3.09010230 & 2.768$\times$10$^{-5}$ & AU \\ 
$q$ & 1.57898175 & 1.365$\times$10$^{-6}$ 
       & 1.57897812 & 4.359$\times$10$^{-6}$ 
       & 1.57897862 & 3.187$\times$10$^{-6}$ & AU \\ 
$i$  & 9.37671494 & 1.640$\times$10$^{-5}$ 
       & 9.37689674 & 8.154$\times$10$^{-5}$ 
       & 9.37689151 & 5.119$\times$10$^{-5}$ &  deg \\ 
$\Omega$& 3.82085037 & 9.844$\times$10$^{-5}$ 
                 & 3.82057683 & 1.816$\times$10$^{-4}$ 
                 & 3.82058910 & 1.183$\times$10$^{-4}$ & deg \\ 
$\omega$ & 152.44185173 & 2.482$\times$10$^{-4}$ 
                 & 152.44464445 & 6.878$\times$10$^{-4}$ 
                 & 152.44467908 & 3.921$\times$10$^{-4}$ & deg \\ 
$M$ \tablenotemark{\dagger} & 3.89813545 & 8.216$\times$10$^{-5}$ 
                                               & 3.89794589 & 2.568$\times$10$^{-4}$ 
                                               & 3.89793566 & 1.425$\times$10$^{-4}$ & deg \\ 
$t_{\mathrm{P}}$ \tablenotemark{\ddagger} & 55482.36069 & 4.543$\times$10$^{-4}$ 
                                                                      & 55482.36498 & 1.251$\times$10$^{-3}$ 
                                                                      & 55482.36503 & 7.022$\times$10$^{-4}$ & TT \\ 
$P$ \tablenotemark{\sharp} & 5.43291573 & 4.529$\times$10$^{-7}$ 
                                            & 5.43209588 & 1.088$\times$10$^{-4}$ 
                                            & 5.43209664 & 7.299$\times$10$^{-5}$&  yr \\ 
$A_{1}$ & -- & -- & 1.5377$\times$10$^{-8}$   & 3.935$\times$10$^{-9}$ 
                           & 1.5750$\times$10$^{-8}$   & 2.477$\times$10$^{-9}$ & AU day$^{-2}$\\ 
$A_{2}$ & -- & -- & 7.1943$\times$10$^{-9}$   & 1.919$\times$10$^{-9}$ 
                           & 7.0860$\times$10$^{-9}$   & 1.216$\times$10$^{-9}$ & AU day$^{-2}$ \\ 

\enddata
\tablenotetext{a}{~Orbit determination without including nongravitational parameters $A_{1}$ and $A_{2}$.}
\tablenotetext{b}{~Orbit determination with $A_{1}$ and $A_{2}$.}
\tablenotetext{c}{~Orbit determination with $A_{1}$ and $A_{2}$ from 2,000 Monte Carlo random exclusion runs.}
\tablenotetext{\dagger}{~Mean anomaly at osculating epoch JD 2455504.347776 = 2010-Nov-03.847776 TT.}
\tablenotetext{\ddagger}{~Time of perihelion passage in modified Julian day (MJD = JD - 2400000.5).}
\tablenotetext{\sharp}{~Orbital period in Julian year.}
\tablecomments{A zero normal nongravitational parameter $A_{3}$ is assumed for solutions NG$_1$ and NG$_2$. The numbers of used astrometric observations of solutions PG and NG$_1$ are respectively 881 and 888 out of 971 available observations in total, which all satisfy the criterion of residuals $\le$2\arcsec.0. RMS of solutions PG and NG$_1$ are $\pm0\arcsec.609$~and $\pm0\arcsec.576$, respectively. There is no mathematically meaningful RMS for solution NG$_2$.}
\end{deluxetable}
\clearpage

\begin{deluxetable}{lcrrrrrc}
\tablecaption{Viewing Geometry of 332P/Ikeya-Murakami in the CFHT and CSS Data
\label{tab_geo}}
\tablewidth{0pt}
\tablehead{ 
\colhead{Date} & \colhead{Survey} & \colhead{$r$}  & \colhead{$\mathit{\Delta}$} & \colhead{$\alpha$\tablenotemark{\dagger}}  & \colhead{$\varepsilon$\tablenotemark{\ddagger}}  & \colhead{$t - t_{\mathrm{P}}$\tablenotemark{\sharp}}   & \colhead{$X$\tablenotemark{\ast}} \\
\colhead{(UT)} & & \colhead{(AU)}  & \colhead{(AU)} & \colhead{(\degr)}  & \colhead{(\degr)}  & \colhead{(day)}  & 
}
\startdata

2003-09-25 10:14-11:32 & CFHT& 4.046 & 3.050 & 2 & 173 & -593 & 1.04-1.07\\
2003-09-27 09:05-11:48 & CFHT & 4.041 & 3.041 & 1 & 175 & -591 & 1.11-1.10\\\cline{1-8}
2005-04-19 03:43-04:07 & CSS & 1.595 & 1.491 & 37 & 75 & -22 & 1.29-1.40\\
2005-04-30 03:12-03:40 & CSS & 1.584 & 1.553 & 37 & 72 & -11 & 1.24-1.34\\

\enddata

\tablenotetext{\dagger}{Phase angle}
\tablenotetext{\ddagger}{Solar elongation}
\tablenotetext{\sharp}{Time to the nearest perihelion passage of the comet.}
\tablenotetext{\ast}{Air mass, dimensionless}

\end{deluxetable}

\clearpage

\begin{deluxetable}{l|rrrrrrrc}
\tablecaption{Small Bodies in Orbits Similar to 332P/Ikeya-Murakami \label{tab_simorb}}
\tablewidth{0pt}
\tablehead{

Object    &  \colhead{$e$}  & \colhead{$a$} & \colhead{$q$} & \colhead{$i$} & \colhead{$\Omega$}
  & \colhead{$\omega$} & \colhead{$T_{\mathrm{J}}$} & \colhead{$D_{\mathrm{SH}}$} \\
   & & \colhead{(AU)} & \colhead{(AU)} & \colhead{(\degr)} & \colhead{(\degr)} & \colhead{(\degr)}  & &
} 

\startdata
2004 CQ$_{49}$ & 0.454 & 2.765 & 1.510 & 8.24 & 11.56 & 155.29 & 3.17 &  0.17 \\
2011 FE$_{73}$ & 0.447 & 3.124 & 1.729 & 14.44 & 17.64 & 134.69 & 3.01 & 0.19 \\
P/2010 B2 (WISE) & 0.481 & 3.106 & 1.612 & 8.94 & 0.85 & 155.94 & 3.01 & 0.04 \\\cline{1-9}
332P/Ikeya-Murakami & 0.490 & 3.086 & 1.573 & 9.39 & 3.78 & 152.45 & 3.01 & -- \\

\enddata

\tablecomments{
The epoch of all the orbits is JD 2457400.5 TT. See Section \ref{sec_search_frag} for details about the orbital criteria for choosing candidates. For better comparison, the orbit of 332P at the same epoch is also presented in the table.
}

\end{deluxetable}







\end{document}